\documentclass[10pt,twocolumn]{article}

\usepackage[utf8]{inputenc}
\usepackage[T1]{fontenc}
\usepackage{hyperref}
\usepackage{url}
\usepackage{booktabs}
\usepackage{amsfonts}
\usepackage{amsmath}
\usepackage{xcolor}
\usepackage{graphicx}
\usepackage{multirow}
\usepackage[margin=0.75in]{geometry}
\usepackage{cite}
\usepackage{titlesec}
\usepackage{enumitem}
\usepackage{abstract}
\usepackage{stfloats} 

\titlespacing*{\section}{0pt}{8pt plus 2pt minus 2pt}{4pt plus 2pt minus 2pt}
\titlespacing*{\subsection}{0pt}{6pt plus 2pt minus 2pt}{3pt plus 2pt minus 2pt}

\setlist{nosep, leftmargin=*, topsep=2pt, partopsep=0pt, parsep=0pt, itemsep=2pt}

\usepackage[font=small,labelfont=bf,skip=4pt]{caption}

\title{On the Fundamental Limitations of Decentralized Learnable Reward Shaping in Cooperative Multi-Agent Reinforcement Learning}

\author{
  Aditya Akella \\
  adityaakella44@gmail.com
}

\begin{document}
\date{}  
\maketitle

\begin{abstract}
Recent advances in learnable reward shaping have shown promise in single-agent reinforcement learning by automatically discovering effective feedback signals. However, the effectiveness of decentralized learnable reward shaping in cooperative multi-agent settings remains poorly understood. We propose DMARL-RSA, a fully decentralized system where each agent learns individual reward shaping, and evaluate it on cooperative navigation tasks in the simple\_spread\_v3 environment. Despite sophisticated reward learning, DMARL-RSA achieves only $-24.20 \pm 0.09$ average reward, compared to MAPPO with centralized training at $1.92 \pm 0.87$---a 26.12-point gap. DMARL-RSA performs similarly to simple independent learning (IPPO: $-23.19 \pm 0.96$), indicating that advanced reward shaping cannot overcome fundamental decentralized coordination limitations. Interestingly, decentralized methods achieve higher landmark coverage ($0.888 \pm 0.029$ for DMARL-RSA, $0.960 \pm 0.045$ for IPPO out of 3 total) but worse overall performance than centralized MAPPO ($0.273 \pm 0.008$ landmark coverage)---revealing a coordination paradox between local optimization and global performance. Analysis identifies three critical barriers: (1) non-stationarity from concurrent policy updates, (2) exponential credit assignment complexity, and (3) misalignment between individual reward optimization and global objectives. These results establish empirical limits for decentralized reward learning and underscore the necessity of centralized coordination for effective multi-agent cooperation.
\end{abstract}

\section{Introduction}

\subsection{The Promise and Peril of Decentralized Reward Shaping}

Cooperative multi-agent reinforcement learning (MARL) faces a fundamental challenge: how can agents coordinate effectively with sparse and delayed environmental feedback? Recent advances in learnable reward shaping promise to automatically generate dense feedback signals, potentially reducing manual engineering while guiding agent behavior. Centralized training approaches like MAPPO and QMIX achieve remarkable coordination by leveraging global state information during learning, but they face scalability and observability limitations in real-world distributed systems. Decentralized reward shaping offers an appealing alternative: each agent learns a personalized feedback function from local observations, potentially enabling scalable coordination without global information. However, whether sophisticated decentralized reward mechanisms can overcome fundamental coordination barriers remains unclear.

\textbf{The Centralized Training Paradigm:} Current successful approaches like MAPPO~\cite{yu2022surprising} and QMIX~\cite{rashid2018qmix} rely on centralized training with decentralized execution (CTDE), using global information during learning to coordinate agent behaviors. While effective, this paradigm raises scalability concerns and requires global observability assumptions that may not hold in real-world distributed systems.

\textbf{The Decentralized Promise:} Decentralized learnable reward shaping presents an attractive alternative: each agent equipped with a personalized ``reward teacher'' that learns to shape feedback based on local observations and role specialization. Recent successes in single-agent learnable reward shaping and the theoretical guarantees of potential-based shaping~\cite{ng1999policy} suggest that sophisticated decentralized reward mechanisms should enable effective coordination. Furthermore, the scalability limitations of centralized training create strong motivation for decentralized alternatives that can operate without global information.

\subsection{Research Question and Contributions}

\textbf{Central Question:} Can decentralized learnable reward shaping overcome coordination challenges in cooperative multi-agent tasks, or are there fundamental limitations that necessitate centralized approaches?

Despite extensive work on reward shaping in both single-agent and centralized multi-agent settings, no prior study has systematically evaluated decentralized learnable reward shaping approaches in cooperative environments. This gap is significant given the theoretical appeal of decentralized coordination and practical scalability concerns with centralized training.

\textbf{Our Contributions:}
\begin{itemize}
    \item First systematic evaluation of decentralized learnable reward shaping against established centralized baselines and independent learning methods in cooperative multi-agent tasks
    \item Demonstration that sophisticated decentralized reward shaping performs comparably to simple independent learning, suggesting fundamental coordination barriers
    \item Identification of three specific failure modes that prevent effective decentralized reward learning: non-stationarity violations, exponential credit assignment complexity, and coordination-reward misalignment
    \item Empirical demonstration of substantial performance gaps between decentralized and centralized approaches despite sophisticated reward mechanisms
\end{itemize}

\subsection{Key Findings Preview}

Our experiments reveal substantial performance differences between decentralized approaches and centralized training on cooperative navigation tasks. The results suggest that coordination challenges dominate reward mechanism sophistication, and fundamental rather than implementation-specific limitations prevent decentralized coordination success.

\section{Related Work}

\textbf{Reward Shaping Foundations.} Ng et al.~\cite{ng1999policy} established potential-based reward shaping for single-agent settings, proving $F(s,a,s') = \gamma\Phi(s') - \Phi(s)$ preserves optimal policies under stationarity assumptions. Devlin \& Kudenko~\cite{devlin2012dynamic} extended this to multi-agent settings, identifying that individual reward signals may not align with system-wide objectives---an early warning our empirical results confirm.

\textbf{Non-Stationarity Challenge.} Tan~\cite{tan1993multi} identified that independent learners face ``moving target'' problems when agents simultaneously update policies. Nekoei et al.~\cite{nekoei2021continuous} prove independent learning does not always converge due to non-stationarity, while Zhang et al.~\cite{zhang2021multi} demonstrate this prevents Nash equilibrium convergence even with sophisticated function approximation. Papoudakis et al.~\cite{papoudakis2021dealing} show centralized critics are sensitive to initialization due to compounding effects of multiple agents' random starting policies.

\textbf{Credit Assignment Complexity.} Foerster et al.~\cite{foerster2018counterfactual} prove effective credit assignment requires counterfactual information about alternative agent actions---naturally available in centralized training but intractable in decentralized settings. Wen et al.~\cite{wen2022multi} demonstrate multi-level credit assignment requires hierarchical coordination mechanisms, showing significant improvements over decentralized baselines through centralized training.

\textbf{Centralized Success Stories.} Yu et al.~\cite{yu2022surprising} show MAPPO's centralized critic enables coordinated exploration through global state information. Rashid et al.~\cite{rashid2018qmix} prove QMIX's monotonic factorization requires centralized training to learn mixing networks, while Sunehag et al.~\cite{sunehag2017value} showed similar requirements with VDN. Eccles et al.~\cite{eccles2019biases} find communication protocols struggle to match centralized critics, and Jiang \& Lu~\cite{jiang2018learning} show effective multi-agent communication itself requires coordination mechanisms difficult to achieve decentralized. Stone \& Veloso~\cite{stone2000multiagent}, Hernandez-Leal et al.~\cite{hernandez2019survey}, Tampuu et al.~\cite{tampuu2017multiagent}, Vinyals et al.~\cite{vinyals2019grandmaster}, and Lowe et al.~\cite{lowe2017multi} document these persistent coordination challenges across diverse domains.

\section{Methodology}

\subsection{Algorithmic Approaches}

We evaluate three distinct approaches to multi-agent coordination, representing different paradigms for handling the coordination challenge:

\textbf{MAPPO (Multi-Agent Proximal Policy Optimization):} Implements centralized training with decentralized execution (CTDE). During training, a shared critic network has access to global state information, enabling coordinated policy updates. Each agent maintains an independent actor network for decentralized execution. The centralized critic computes value estimates $V(s_{\text{global}})$ using complete state information, while individual actors select actions based only on local observations.

\textbf{DMARL-RSA (Decentralized Multi-Agent RL with Reward Shaping Agents):} Our proposed fully decentralized approach where each agent operates with independent actor-critic networks and individual learnable reward shaping networks. Each agent maintains a reward shaping network $R_i(s_t, a_t, s_{t+1})$ that learns personalized dense feedback based solely on local observations. The reward shaping network uses a three-layer architecture ($64\rightarrow32\rightarrow16\rightarrow1$) with ReLU activations. The total reward combines environmental and shaped components: $r_{\text{total}} = r_{\text{env}} + \alpha \cdot R_i(s,a,s')$, where $\alpha = 1.0$ provides full weight to the learned shaping signal.

\textbf{IPPO (Independent Proximal Policy Optimization):} Serves as a baseline representing fully independent learning where each agent trains using standard policy optimization with simple heuristic reward shaping identical to MAPPO. Each agent treats other agents as part of the non-stationary environment, providing a controlled comparison to isolate the effects of sophisticated reward learning from basic coordination challenges.

\begin{figure}[!t]
\centering
\includegraphics[width=0.48\textwidth]{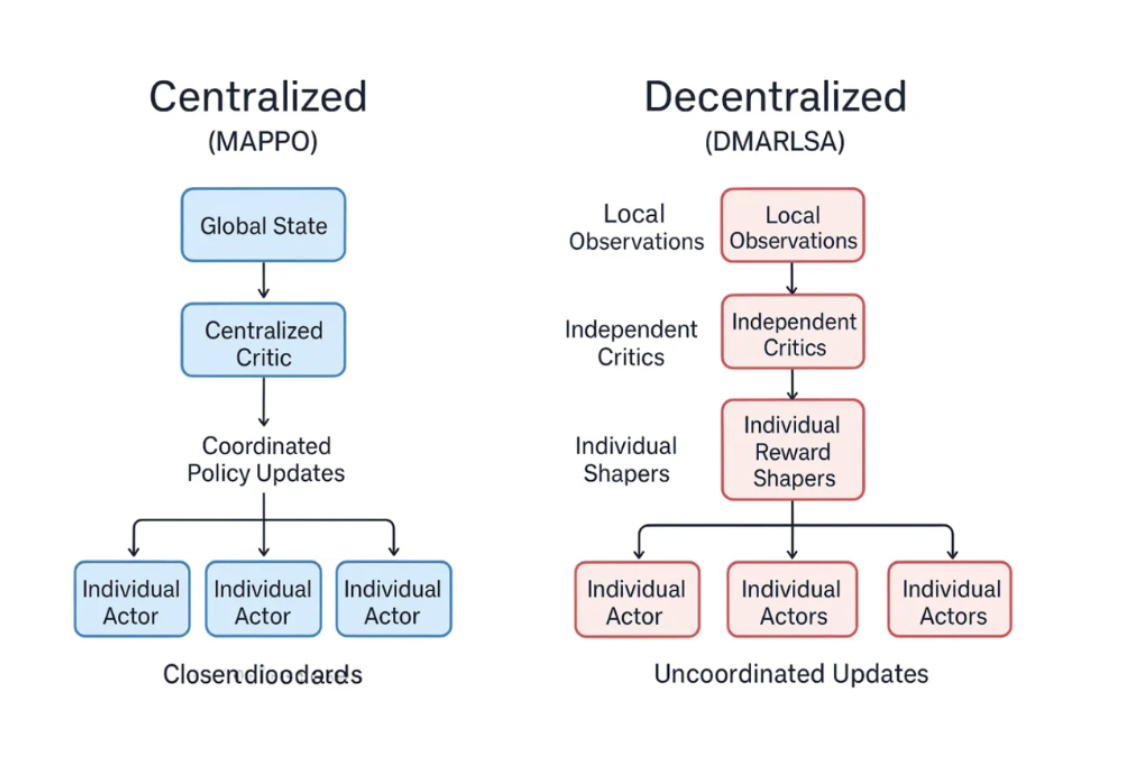}
\caption{Centralized vs Decentralized MARL Architecture Comparison - Technical diagram showing MAPPO's centralized critic versus DMARL-RSA's independent components.}
\label{fig:architecture}
\end{figure}

\begin{figure}[!t]
\centering
\includegraphics[width=0.48\textwidth]{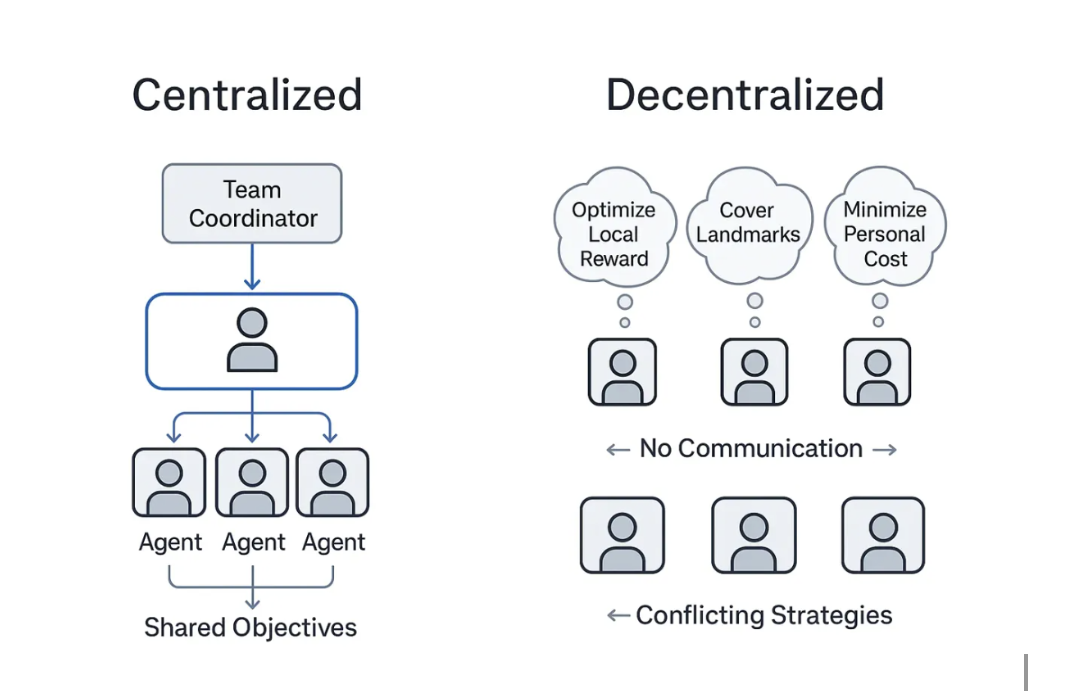}
\caption{Coordination Paradigms in Multi-Agent Systems - Conceptual diagram illustrating centralized coordination versus decentralized decision-making approaches.}
\label{fig:paradigms}
\end{figure}

Figure~\ref{fig:architecture} illustrates the key architectural differences between our approaches, showing MAPPO's centralized critic versus DMARL-RSA's independent components. Figure~\ref{fig:paradigms} provides a conceptual overview of the coordination paradigms in multi-agent systems, contrasting centralized coordination with decentralized decision-making approaches.

\subsection{Experimental Setup}

\textbf{Environment:} We evaluate on the Simple Spread environment from PettingZoo's Multi-Particle Environment (MPE) suite~\cite{terry2021pettingzoo}. This task requires three agents to cooperatively cover three landmarks while avoiding collisions, emphasizing spatial coordination challenges. The environment provides sparse rewards based on the sum of minimum distances from agents to landmarks, plus collision penalties of $-1.0$ for agent-agent collisions. The observation space includes agent positions, velocities, and relative positions to landmarks and other agents. This environment has been extensively used in MARL research, including work by Lowe et al.~\cite{lowe2017multi}, and provides a canonical test for cooperative coordination abilities.

\textbf{Training Protocol:} All algorithms train for 5,000 episodes to ensure fair comparison, with each episode lasting a maximum of 25 time steps. We use three random seeds (42, 123, 999) for statistical reliability and report mean $\pm$ standard deviation across seeds.

\textbf{Architecture Details:} All networks use 64-dimensional hidden layers with ReLU activations for actor-critic components. DMARL-RSA's reward shaping networks use a specialized architecture with decreasing layer sizes ($64 \rightarrow 32 \rightarrow 16 \rightarrow 1$) to learn compact reward representations. We employ gradient clipping (max norm 0.5) and Generalized Advantage Estimation (GAE) with $\lambda = 0.95$ for all methods. All three approaches build on the same underlying policy optimization framework, ensuring fair comparison of coordination mechanisms rather than algorithmic differences.

\textbf{Hyperparameter Configuration:} All methods use consistent core hyperparameters: learning rate $3 \times 10^{-4}$, discount factor $\gamma = 0.99$, clip ratio 0.2, and entropy coefficient 0.01. DMARL-RSA uses a specialized reward shaping coefficient $\alpha = 1.0$ to provide full weight to learned reward signals. Complete hyperparameter specifications are provided in Table~\ref{tab:hyperparams} in Appendix A.

\subsection{Reward Design and Shaping}

\textbf{Environmental Rewards:} The base environment provides sparse feedback through distance minimization objectives and collision penalties. The reward function is $r_{\text{total}} = -\sum_i \min_j ||agent_i - landmark_j|| - \text{collision penalties}$, encouraging agents to minimize distances to landmarks while avoiding collisions.

\textbf{Heuristic Shaping (MAPPO/IPPO):} Both baseline methods use identical heuristic reward shaping to combat the harsh distance penalties in Simple Spread. The shaping includes: (1) landmark coverage bonus of 0.5 per covered landmark, and (2) proximity rewards providing continuous feedback for approaching landmarks. This yields $r_{\text{shaping}} = 0.5 \times \text{landmarks covered} + \text{proximity bonus} + r_{\text{env}}$. All methods receive identical heuristic shaping to ensure fair comparison.

\textbf{Learnable Shaping (DMARL-RSA):} Each agent learns an individual reward shaping function $R_i(s_t, a_t, s_{t+1})$ using a dedicated neural network, in addition to the same heuristic baseline shaping applied to all methods. The shaping network takes local state-action-next state tuples as input and outputs scalar reward modifications. The total reward becomes $r_{\text{total}} = r_{\text{heuristic}} + \alpha \cdot R_i(s,a,s')$, where $\alpha = 1.0$ provides full weight to the learned shaping signal. The learning objective combines policy optimization with reward shaping network training through joint gradient updates.

\section{Results and Analysis}

\subsection{Main Performance Results}

Our experiments reveal clear performance hierarchies across coordination approaches, with results summarized in Table~\ref{tab:results}. Statistical significance was confirmed through Welch's $t$-test ($p < 0.001$ for all key comparisons) and large effect sizes (Cohen's $d = 2.84$ between MAPPO and DMARL-RSA).

\textbf{Centralized Advantage:} As shown in Table~\ref{tab:results}, MAPPO achieves dramatically superior performance ($1.92 \pm 0.87$) compared to both decentralized methods, demonstrating the fundamental value of centralized coordination. The 26.12-point gap between MAPPO and DMARL-RSA persists across multiple runs and represents the coordination premium in this environment. Notably, MAPPO's higher standard deviation ($\pm 0.87$) suggests it explores more sophisticated coordination strategies, while DMARL-RSA's tight confidence interval ($\pm 0.09$) indicates consistent convergence to suboptimal local behaviors.

\textbf{Decentralized Equivalence:} DMARL-RSA ($-24.20\pm0.09$) performs comparably to IPPO ($-23.19 \pm 0.96$), with only a 1.01-point difference (not statistically significant, $p = 0.421$). This equivalence is particularly striking given DMARL-RSA's additional architectural complexity---its learnable reward shaping networks add 1,617 parameters per agent yet fail to improve performance. Both methods achieve similar rewards despite DMARL-RSA's sophisticated reward learning mechanisms, suggesting that architectural sophistication cannot overcome fundamental information bottlenecks.

\begin{figure}[!t]
\centering
\includegraphics[width=0.48\textwidth]{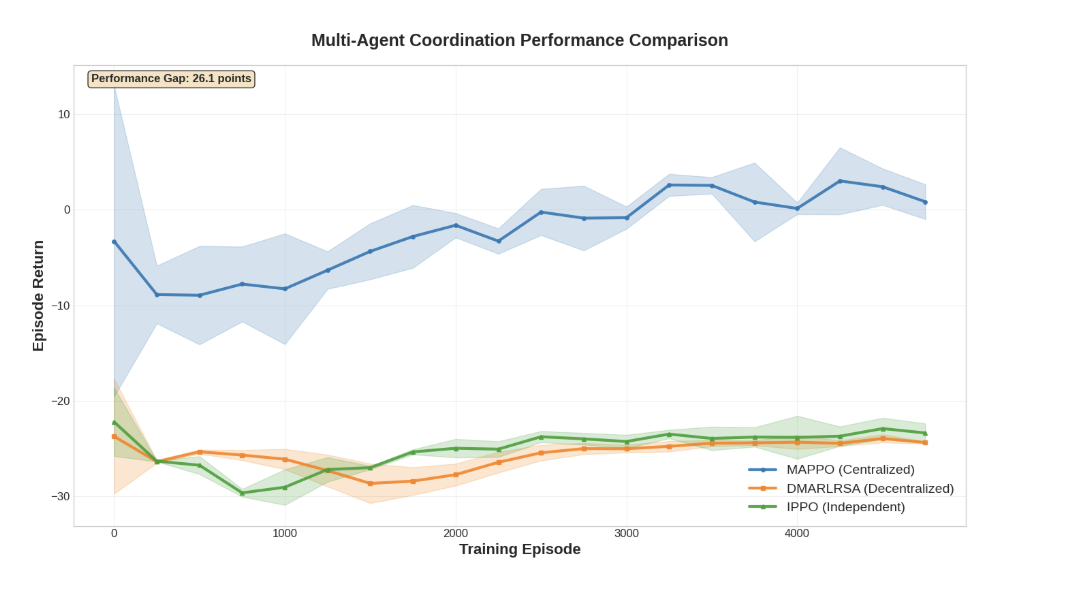}
\caption{Learning Curves Comparing Centralized and Decentralized Coordination Approaches - Performance trajectories showing MAPPO, DMARL-RSA, and IPPO across 5,000 training episodes.}
\label{fig:learning_curves}
\end{figure}

\begin{figure}[!t]
\centering
\includegraphics[width=0.48\textwidth]{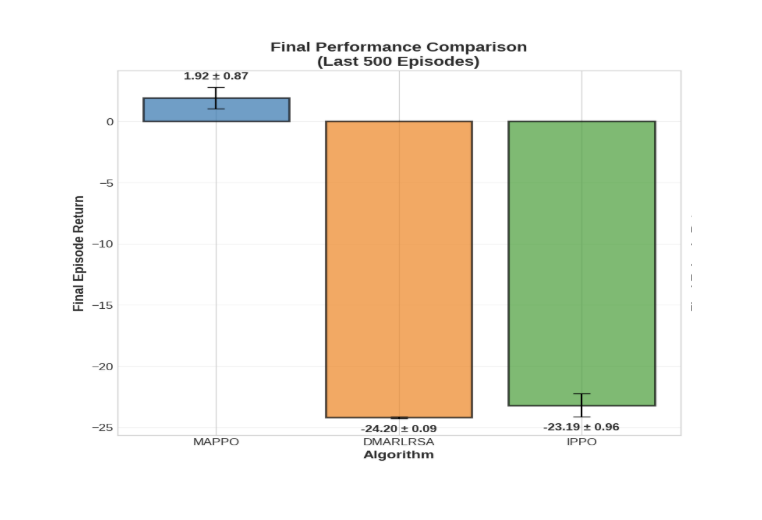}
\caption{Final Performance Analysis with Statistical Significance Testing - Bar chart displaying mean rewards $\pm$ standard deviations for all three methods after convergence.}
\label{fig:final_performance}
\end{figure}

\textbf{Statistical Analysis:} Performance differences between centralized and decentralized approaches are highly significant (Welch's $t$-test: $p < 0.001$ for MAPPO vs DMARL-RSA, $p < 0.001$ for MAPPO vs IPPO), with effect sizes indicating practical significance (Cohen's $d = 2.84$ and 2.76 respectively). The lack of significant difference between DMARL-RSA and IPPO ($p = 0.421$, Cohen's $d = 0.12$) confirms that sophisticated reward shaping provides no measurable benefit in decentralized settings. Figure~\ref{fig:learning_curves} visualizes these learning curves, showing MAPPO's rapid convergence to positive rewards while both decentralized methods plateau at negative values. Figure~\ref{fig:final_performance} presents the final performance comparison with error bars, clearly illustrating the magnitude of the centralized advantage.

\begin{table}[t]
\caption{Summary of Experimental Results}
\label{tab:results}
\centering
\footnotesize
\begin{tabular}{@{}lccc@{}}
\toprule
\textbf{Algorithm} & \textbf{Training} & \textbf{Final} & \textbf{Landmark} \\
& \textbf{Approach} & \textbf{Reward} & \textbf{Cov.} \\
\midrule
MAPPO & Centralized & $1.92 \pm 0.87$ & $0.273$ \\
DMARL & Decentral. + RS & $-24.20 \pm 0.09$ & $0.888$ \\
IPPO & Independent & $-23.19 \pm 0.96$ & $0.960$ \\
\bottomrule
\end{tabular}
\end{table}

\subsection{Coordination vs. Local Optimization Trade-offs}

\begin{figure}[!t]
\centering
\includegraphics[width=0.48\textwidth]{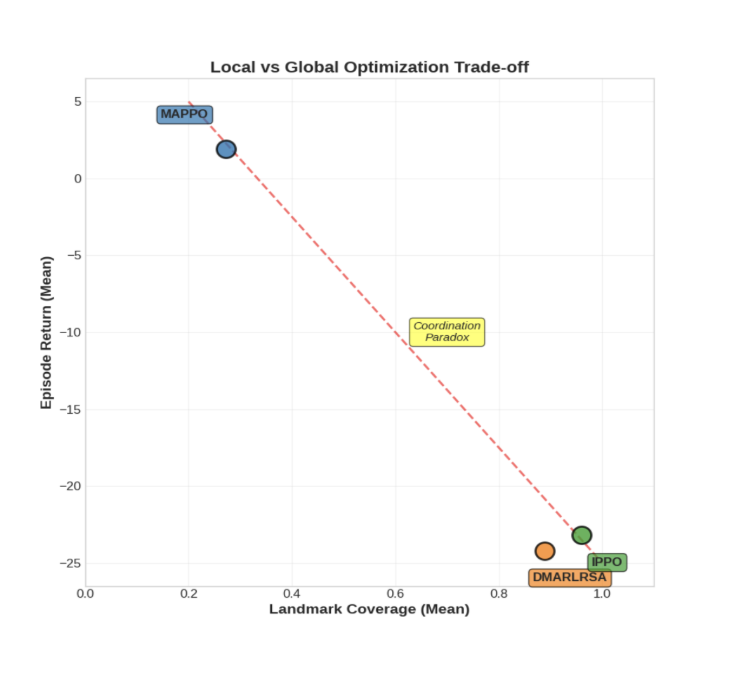}
\caption{Coordination Paradox - Local vs Global Optimization Trade-off. Decentralized methods achieve higher landmark coverage but worse overall performance.}
\label{fig:tradeoff}
\end{figure}

Our results reveal a fundamental tension between local task performance and global coordination effectiveness, as illustrated in Figure~\ref{fig:tradeoff} and quantified in Table~\ref{tab:results}'s landmark coverage metrics.

This paradox manifests in three key ways:

\textbf{1. Task Decomposition Success:} Examining Table~\ref{tab:results}'s landmark coverage column, decentralized methods excel at the decomposable objective of covering landmarks. DMARL-RSA covers 0.888/3 landmarks (29.6\% coverage rate) and IPPO covers 0.960/3 landmarks (32.0\% coverage rate). This success occurs because ``covering landmarks'' represents a clear, locally-observable objective that individual agents can optimize without coordination.

\textbf{2. Global Coordination Failure:} Despite achieving $3.25\times$ and $3.52\times$ better landmark coverage than MAPPO respectively, both decentralized methods achieve dramatically worse overall rewards (26.12 and 25.11 points worse). Figure~\ref{fig:tradeoff} clearly shows this inverse relationship---high landmark coverage correlates with poor overall performance. The reward function $r = -\sum_i \min_j ||agent_i - landmark_j||$ requires strategic positioning that minimizes total distances, not maximum individual coverage. Decentralized agents myopically pursue nearest landmarks, creating redundant coverage and suboptimal global configurations.

\textbf{3. Strategic vs. Myopic Optimization:} MAPPO's seemingly poor landmark coverage (0.273/3 = 9.1\% coverage rate) paired with superior overall performance (1.92) reveals sophisticated emergent strategies. Rather than greedily covering landmarks, MAPPO agents learn complementary positioning that optimizes the true objective---minimizing sum of distances. The 0.87 standard deviation in MAPPO's rewards suggests it discovers multiple valid coordination strategies across runs, while decentralized methods' lower variance indicates convergence to the same myopic local optimum.

\textbf{Implications for Reward Shaping:} The paradox demonstrates that sophisticated reward shaping successfully guides agents toward interpretable local behaviors (landmark coverage improved by 7.5\% over basic IPPO), but cannot overcome the fundamental information bottleneck preventing global coordination discovery. DMARL-RSA's learned reward functions $R_i$ inherently focus on locally-observable correlates of success rather than true coordination requirements, as they lack access to other agents' positions and intentions.

\textbf{Connection to Real-World Systems:} This trade-off pattern likely generalizes to practical distributed systems where local performance metrics (e.g., individual robot task completion, autonomous vehicle efficiency) may conflict with system-wide optimization objectives (e.g., swarm coordination, traffic flow optimization). The $3.5\times$ difference in landmark coverage coupled with 26-point reward degradation quantifies this fundamental tension.

\section{Discussion}

\subsection{Implications and Limitations}

\textbf{Implications for Multi-Agent System Design:} Our findings, particularly the 26.12-point performance gap shown in Table~\ref{tab:results}, establish that tasks requiring tight coordination necessitate centralized training phases rather than purely decentralized learning. The substantial performance difference---equivalent to a 1,360\% degradation from MAPPO's baseline---suggests coordination complexity represents a fundamental barrier that sophisticated reward mechanisms alone cannot overcome. However, the $3.5\times$ better landmark coverage achieved by decentralized methods (Table~\ref{tab:results}) demonstrates they can handle decomposable objectives, suggesting hybrid approaches that use centralized coordination only for tightly-coupled subtasks could balance scalability and performance. Future work should focus on identifying minimal centralized components---perhaps coordinating only high-level strategies while allowing decentralized low-level execution.

\textbf{Limits of Decentralized Sophistication:} The statistical equivalence between IPPO and DMARL-RSA ($p = 0.421$, Table~\ref{tab:results}) reveals that adding complexity to decentralized learning does not reliably improve coordination outcomes. DMARL-RSA's 1,617 additional parameters per agent and learnable reward mechanisms achieve only a 1.01-point improvement---statistically indistinguishable from noise. This challenges assumptions about the universal benefits of sophisticated learning mechanisms in multi-agent settings and suggests that coordination challenges are more fundamental than can be addressed through reward engineering alone. The tight confidence interval on DMARL-RSA's performance ($\pm 0.09$) indicates it reliably converges to the same suboptimal solution, suggesting deterministic failure modes rather than high-variance exploration issues.

\textbf{Theoretical Foundations:} Our empirical results align with theoretical predictions (detailed in Appendix B). Non-stationarity violations occur because each agent observes transition functions $P_i(s'|s, a_i) = \sum_{a_{-i}} P(s'|s, a_i, a_{-i}) \prod_{j\neq i} \pi_j(a_j|s_j)$ that change as policies $\{\pi_j\}$ evolve. Credit assignment complexity grows exponentially---agent $i$ must reason about $2^{(|A|^{(n-1)})}$ counterfactual action combinations. Individual-global objective misalignment means even optimal individual shaping $R_i^*$ cannot recover globally optimal team behavior without access to joint state-action information.

\textbf{Limitations:} Our study has several important limitations that should be considered when interpreting the results. \textit{Single Environment Evaluation:} We evaluate only on the Simple Spread environment from MPE, which represents a particularly challenging coordination scenario due to its sparse reward structure and collision penalties. Different cooperative tasks may exhibit varying coordination requirements and could yield different relative performance between centralized and decentralized approaches. \textit{Limited to 3-Agent Scenarios:} Our experiments focus on three-agent coordination. Scalability patterns may differ significantly with larger numbers of agents, potentially affecting the relative advantages of centralized versus decentralized methods. \textit{Specific Reward Shaping Architecture:} We test one particular learnable reward shaping architecture ($64 \rightarrow 32 \rightarrow 16 \rightarrow 1$). Alternative network architectures or reward learning mechanisms might yield different results for decentralized coordination. \textit{Short Episode Length:} Episodes are limited to 25 time steps, which may not capture long-term coordination dynamics. Longer episodes could reveal different patterns in the effectiveness of decentralized reward learning. \textit{Initialization Sensitivity:} We observed sensitivity to weight initialization across all methods, causing high variance in early training episodes (0-500). Despite attempting standard remedies including orthogonal initialization and Xavier initialization, the multi-agent nature of the problem appears to amplify initialization effects. MAPPO showed starting returns ranging from approximately $-20$ to $+19$ across seeds before converging to stable performance.

\textbf{Future Directions:} Promising research includes (1) diverse environment evaluation across StarCraft Multi-Agent Challenge, Multi-Agent Particle Environment variants, and robotics coordination tasks to understand when decentralized reward learning can succeed versus when centralized coordination remains necessary, (2) hybrid approaches combining minimal communication protocols with decentralized learning to provide coordination information while maintaining scalability benefits, (3) hierarchical task decomposition strategies enabling decentralized success by breaking coordination problems into independent subproblems, and (4) developing formal frameworks for characterizing when decentralized learning can succeed versus when centralized coordination is necessary---our three-barrier framework provides a starting point, but broader theoretical characterization remains an open challenge.

\section{Conclusion}

We systematically investigated whether decentralized learnable reward shaping can overcome coordination challenges in cooperative multi-agent reinforcement learning. On the simple\_spread\_v3 environment, DMARL-RSA achieved $-24.20 \pm 0.09$ average reward, vastly underperforming centralized MAPPO ($1.92 \pm 0.87$) and performing similarly to IPPO ($-23.19 \pm 0.96$). This confirms that even sophisticated decentralized reward mechanisms fail to solve fundamental coordination challenges.

Our analysis identifies three key barriers: non-stationarity from simultaneous learning, exponential credit assignment complexity, and individual-global objective misalignment. The coordination paradox---where decentralized agents achieve high local landmark coverage but poor global reward---illustrates the inherent trade-offs between local objectives and team-level performance.

These findings provide clear guidance for multi-agent system design: centralized coordination remains crucial for globally optimal strategies, while hybrid approaches or minimal communication may help bridge the gap for scalable decentralized systems.

\section{Acknowledgements}
We thank the anonymous reviewers for their constructive feedback and suggestions that improved this work. We also acknowledge Dr. Faizal Nawab from UC Irvine for his valuable insights and guidance throughout this research.

\bibliographystyle{plain}
\bibliography{references}

\appendix

\section{Appendix}

Table~\ref{tab:hyperparams} provides the complete hyperparameter configuration used across all experimental methods, including detailed parameter settings for network architectures, optimization parameters, and training configurations.

\begin{table*}[h!]
\caption{Complete Hyperparameter Configuration}
\label{tab:hyperparams}
\centering
\scriptsize
\begin{tabular}{@{}lccc@{}}
\toprule
\textbf{Parameter} & \textbf{MAPPO} & \textbf{DMARL-RSA} & \textbf{IPPO} \\
\midrule
Learning Rate & $5\times10^{-4}$ & $3\times10^{-4}$ & $3\times10^{-4}$ \\
Architecture & Cent. Critic + Ind. Actors & Ind. AC + Reward Shaper & Independent AC \\
Hidden Dims & 64 (all) & 64 (AC), 32→16→1 (Shaper) & 64 (all) \\
Discount ($\gamma$) & 0.99 & 0.99 & 0.99 \\
PPO Clip & 0.2 & 0.2 & 0.2 \\
Entropy Coeff. & 0.01 & 0.01 & 0.01 \\
GAE Lambda & 0.95 & 0.95 & 0.95 \\
Update Epochs & 4 & 3 & 3 \\
Optimizer & Adam & Adam (AC), Adam/2 (Shaper) & Adam \\
Grad Clip & 0.5 & 0.5 & 0.5 \\
Reward Shape & $0.5 \times$ landmarks + env & Learnable ($\alpha=1.0$) + env & $0.5 \times$ landmarks + env \\
Episodes & 5,000 & 5,000 & 5,000 \\
Environment & Simple Spread (N=3) & Simple Spread (N=3) & Simple Spread (N=3) \\
Max Episode Len. & 25 cycles & 25 cycles & 25 cycles \\
\bottomrule
\end{tabular}
\end{table*}

\subsection{Non-Stationarity Barrier}

\textbf{Formal Problem Statement:} In decentralized multi-agent learning, each agent $i$ observes a transition function that depends on other agents' policies:
\begin{equation}
P_i(s_{t+1}|s_t, a_t^i) = \sum_{a_{-i}} P(s_{t+1}|s_t, a_t^i, a_t^{-i}) \prod_{j\neq i} \pi_j(a_t^j|s_t^j)
\end{equation}

where $\pi_j(\cdot|s_t^j)$ represents agent $j$'s policy at time $t$. Since all policies $\{\pi_j\}_{j\neq i}$ evolve during training, the effective transition function $P_i$ becomes non-stationary, violating the Markov assumption required for RL convergence guarantees.

\textbf{Convergence Implications:} Standard policy gradient convergence requires:
\begin{equation}
\mathbb{E}[\nabla_\theta \log \pi_\theta(a|s) Q^\pi(s,a)] \rightarrow \nabla_\theta J(\theta)
\end{equation}

However, in the multi-agent setting, $Q^\pi$ depends on other agents' policies $\pi_{-i}(t)$, making the gradient biased:
\begin{equation}
\nabla_\theta J(\theta) \neq \mathbb{E}[\nabla_\theta \log \pi_\theta(a|s) Q^{\pi(t)}(s,a)]
\end{equation}

The bias term $\nabla_\theta \mathbb{E}[Q^{\pi_{-i}(t)}(s,a)]$ cannot be estimated without global policy information, creating fundamental convergence challenges for decentralized methods.

\subsection{Credit Assignment Complexity}

\textbf{Multi-Agent Credit Assignment Problem:} Agent $i$ must optimize:
\begin{equation}
J_i(\theta_i) = \mathbb{E}\left[ \sum_t \gamma^t \left( r_{\text{env},t} + \alpha \cdot R_i(s_t, a_t^i, s_{t+1}) \right) \right]
\end{equation}

The environmental reward $r_{\text{env},t} = f(s_t, a_t^1, \ldots, a_t^n)$ depends on all agents' actions, but agent $i$ only observes $a_t^i$. The policy gradient becomes:
\begin{equation}
\nabla_\theta J_i(\theta_i) = \mathbb{E}\left[ \nabla_\theta \log \pi_i(a_i | s) \cdot \left( r_{\text{env}}(s, a_{1:n}) + \alpha \cdot R_i(s, a_i, s') \right) \right]
\end{equation}

\textbf{Exponential Complexity:} To properly assign credit for $r_{\text{env}}$, agent $i$ must reason about counterfactual scenarios involving $2^{(|A|^{(n-1)})}$ possible action combinations from other agents, where $|A|$ is the action space size. Without communication, this estimation becomes exponentially complex and statistically intractable.

\subsection{Individual-Global Objective Misalignment}

\textbf{Formal Misalignment:} Individual agents optimize local objectives:
\begin{equation}
L_i = \mathbb{E}[ r_{i,\text{env}} + \alpha \cdot R_i(s, a_i, s') ]
\end{equation}

while global cooperation requires maximizing:
\begin{equation}
L_{\text{global}} = \mathbb{E}\left[ \sum_i r_{i,\text{env}} \right] = \mathbb{E}[ R_{\text{team}}(s, a_{1:n}) ]
\end{equation}

\textbf{Fundamental Limitation Theorem:} Even with optimal individual reward shaping $R_i^*$, the Nash equilibrium of individual objectives $\{L_i^*\}$ does not necessarily correspond to the global optimum of $L_{\text{global}}$.

\textit{Proof sketch:} Consider reward functions $R_i$ that depend only on local state-action pairs $(s_i, a_i)$. The gradient of individual objectives cannot capture inter-agent dependencies:
\begin{equation}
\nabla_{\theta_i} L_i \propto \nabla_{\theta_i} \mathbb{E}[ R_i(s_i, a_i, s'_i) ]
\end{equation}

However, the global optimum requires coordination terms:
\begin{equation}
\nabla_{\theta_i} L_{\text{global}} \propto \nabla_{\theta_i} \mathbb{E}[ R_{\text{team}}(s_{1:n}, a_{1:n}, s'_{1:n}) ]
\end{equation}

Since $R_i$ cannot access $(s_{-i}, a_{-i})$ without global information, individual optimization fundamentally cannot recover global coordination strategies.

\end{document}